\documentstyle[preprint,revtex]{aps}
\tightenlines
\begin{document}
\draft
\begin{title}
{\bf General Non-Static Spherically Symmetric Solutions\\
of Einstein Vacuum Field Equations with $\Lambda$}
\end{title}
\author{\bf Soheila Gharanfoli and Amir H. Abbassi}
\begin{instit}
Department of Physics, School of Sciences,\\
Tarbiat Modarres University, P.O.Box 14155-4838,
Tehran, I.R.Iran.
\end{instit}
\author{\it E-mail: ahabbasi@net1cs.modares.ac.ir}
\begin{abstract}
\begin{center}
{\bf ABSTRACT}
\end{center}
$1-$ It is shown that the upper bound for $\alpha $ in the general solutions
of spherically symmetric vacuum field equations (gr-qc/9812081, $\Lambda =0$)
is nearly $10^3$. This has been obtained by comparing the theoretical
prediction for bending of light and precession of perihelia with observation.
For a significant range of possible values of $\alpha (\alpha >2)$ the metric
is free of coordinate singularity.
$2-$ It is checked that the singularity in the non-static spherically symmetric
solution of Einstein field equations with $\Lambda $ (JHEP 04 (1999) 011,
$\alpha =0$) at the origin is intrinsic.
$3-$ Using the techniques of these two works, a general class of non-static
solutions is presented. They are smooth and finite everywhere and have an
extension larger than the Schwarzschild metric.
$4-$ The geodesic equations of a freely falling material particle for the
general case are solved which reveal a Schwarzschild-de Sitter type potential
field.\\
\end{abstract}


\newpage
\section{\bf Introduction}

According to Birkhoff's theorem the metric for a vacuum spherically symmetric
gravitational field with $\Lambda =0$, has a unique Schwarzschild form
\begin{equation}
ds^2=(1-\frac{2M}{r})dt^2-(1-\frac{2M}{r})^{-1}dr^2
-r^2({d\theta}^2+\sin^2\theta d\phi^2)\;\label{eq:sch}
\end{equation}
and with $\Lambda\neq0$ it has a unique Schwarzschild-de Sitter form
\begin{equation}
ds^2=(1-\frac{2M}{r}-\frac{\Lambda}{3}r^2)dt^2-(1-\frac{2M}{r}
-\frac{\Lambda}{3}r^2)^{-1}
dr^2-r^2(d{\theta}^2+\sin^2\theta d\phi^2)\;\label{eq:scds}
\end{equation}
Eq.~(\ref{eq:sch}) has a coordinate type singularity at $r=2M$ and an intrinsic 
singularity at $r=0$ while Eq.~(\ref{eq:scds}) has two coordinate singularities
at $r\approx 2M$ and $r\approx(\frac{3}{\Lambda})^{\frac{1}{2}}$, and an
intrinsic singularity at $r=0$. The intrinsic singularity is irremovable and
this is indicated by diverging the Riemann tensor scalar invariant
\cite{inver92}
\begin{equation}
{\cal R}^a\;_{bcd}{\cal R}_a\;^{bcd}=\frac{48M^2}{r^6}\;\label{eq:cur}
\end{equation}

For case $\Lambda=0$, a general class of solutions has been obtained which
has the form \cite{abbas98}
\begin{equation}
ds^2=(1-{2M\over r+\alpha M})dt^2-(1-{2M\over r+\alpha M})^{-1}dr^2
-(r+\alpha M)^2(d{\theta}^2+\sin^2\theta d\phi^2)\; \label{eq:alph}
\end{equation}
where $\alpha$ is an arbitrary constant. In Sec.II we primarily discuss the
apparent objection that they are subspaces of the Schwarzschild metric. Though
it seems only a linear change of variables, but $\alpha $ does affect the
curvature of space-time. Then it is shown that staticness is not a necessary
initial condition if $g_{\theta \theta }$ is independent of time. Subsequently
by deriving the equation of precession of perihelia and bending of light in a
gravitational field and comparing them with observational measurements, we
conclude that $\alpha $ may take small as well as large values up to $10^3$.
Even though this may solve the singularity problem, but recent observations
of type Ia supernovae indicate the existence of a positive cosmological
constant \cite{pococo98}. From other side, it has been shown that in the
presence of cosmological constant, using static reference coordinate, is not
suitable \cite{abbkh98}.

For case $\Lambda \neq0$, a non-static solution for this system has been
proposed which has the following form \cite{abbas99}
\begin{eqnarray}
ds^2&=& ({{\sqrt{(1-\frac{2M}{\rho}-\frac{\Lambda}{3}\rho^2)^2
+\frac{4\Lambda}{3}\rho^2}}+(1-\frac{2M}{\rho}-\frac{\Lambda}{3}\rho^2)\over
2})dt^2\nonumber\\
&&- e^{2{\sqrt{\frac{\Lambda}{3}}}t} \:\biggl[({{\sqrt{(1-\frac{2M}{\rho}-
\frac{\Lambda}{3}\rho^2)^2
+\frac{4\Lambda}{3}\rho^2}}
+(1-\frac{2M}{\rho}-\frac{\Lambda}{3}\rho^2)\over
2})^{-1} dr^2 \nonumber\\
& & \;\;\;\;\;\;\;\;\;\;\;\;\;\;\;\;\;\;\;\;\;\;\;\;\;
\;\;\;\;\;\;\;\;\;\;\;\;\;\;\;\;\;\;\;\;\;\;\;\;\;\;
\;\;\;\;\;\;\;\;\;\;\;\;\;\;\;\; {}+
r^2(d{\theta}^2+\sin^2\theta d\phi^2)\biggr]\;
\end{eqnarray}
where $\rho=re^{{\sqrt{\frac{\Lambda}{3}}}t}$. The result which evidently is
free of any singularity for $r\neq0$, is singular at $r=0$. In Sec.III we
show that it is indeed an intrinsic singularity. In Sec.IV, by making use
of the presented techniques in \cite{abbas98,abbas99}, a general class of
non-static solutions will be obtained which has the form
\begin{eqnarray}
ds^2&=& ({{\sqrt{(1-\frac{2M}{\rho}-\frac{\Lambda}{3}\rho^2)^2
+\frac{4\Lambda}{3}\rho^2}}+(1-\frac{2M}{\rho}-\frac{\Lambda}{3}\rho^2)\over
2})dt^2\nonumber\\
&&- e^{2{\sqrt{\frac{\Lambda}{3}}}t} \:\biggl[({{\sqrt{(1-\frac{2M}{\rho}-\frac{\Lambda}{3}\rho^2)^2
+\frac{4\Lambda}{3}\rho^2}}
+(1-\frac{2M}{\rho}-\frac{\Lambda}{3}\rho^2)\over
2})^{-1} dr^2 \nonumber\\
& & \;\;\;\;\;\;\;\;\;\;\;\;\;\;\;\;\;\;\;\;\;\;\;\;\;
\;\;\;\;\;\;\;\;\;\;\;\;\;\;\;\;\;\;\;\;\;\;\;\;\;\;
\;\;\;\; {}+
(r+\alpha M)^2(d{\theta}^2+\sin^2\theta d\phi^2)\biggr]\;
\end{eqnarray}
where here $\rho =(r+\alpha M)e^{{\sqrt{\frac{\Lambda }{3}}}t}$.
These solutions are smooth and finite everywhere even at $r=0$. We show that
they have an extension larger than the Schwarzschild metric. Obviously they
should be checked for completeness before we may call them non-singular. Sec.V
deals with solving the geodesic equations for a freely falling material
particle in the general case and results a potential field which though is
very large at $r\approx 0$ but it is finite. Finally, derivations of
Eq.(\ref{eq:redef}), Eq.(\ref{eq:defcl}) and Eq.(\ref{eq:preth}) which
being used in Sec.II, are presented in Appendixes \ref{sec:deflec} and
\ref{sec:preces}.\\

\section{\bf Case\ ${\bf \Lambda =0}$ , ${\bf \alpha \neq0}$}

Since Eq.~(\ref{eq:alph}) transforms to Eq.~(\ref{eq:sch}) by simply replacing
$r'=r+\alpha M$ with the range of $r'\geq \alpha M$, this may cause a confusion
that ~(\ref{eq:alph}) is a subspace of ~(\ref{eq:sch}). The proof of
completeness usually for a pseudo-Riemannian manifold is not an easy task. We
will show the flaw in this argument by a Riemann counter-example. Taking $R^2$
as a two-space of all points with coordinate $r,\theta $ such that the metric is
\begin{equation}
ds^2=dr^2+r^2 {d\theta}^2\;
\end{equation}
where $\theta =0$ is identified with $\theta =2\pi $ and the point $r=0$ is
included. This plane is complete and non-singular. Also $R'^2$ is a two-space
of all points with coordinate $r,\theta $ such that the metric is
\begin{equation}
ds^2=dr^2+(r+a)^2 {d\theta}^2\; \label{eq:tsp}
\end{equation}
where the range of $r$ and $\theta $ is the same as $R^2$. If we transform
$r'=r+a$ Eq.~(\ref{eq:tsp}) gives
\begin{equation}
ds^2=dr'^2+r'^2 {d\theta }^2 \;, \;\;\;\;\; \mbox{and} \;\;\; a\leq r'\;
\end{equation}
which apparently this means $R'^2\subset R^2$. We show that indeed this is not
valid. Let's consider a subspace of $R^2$ and $R'^2$ by restricting $r<\rho $.
The surface area of $R^2(r<\rho )$ is $\pi \rho^2$ while the surface area of
$R'^2(r<\rho )$ is $\pi (\rho^2 +2a\rho )$. This means for finite $\rho $ we
always have $R^2(r<\rho )\subset R'^2(r<\rho )$. If we take the limit
$\rho \to \infty $ then we get $R^2(r<\rho ) \to R^2$ and
$R'^2(r<\rho ) \to R'^2$. Since $R^2$ is complete and from other side
$R^2(r<\rho )\subset R'^2(r<\rho )$, there is no way except to conclude
that $R^2=R'^2$. This counter-example shows that how the conclusion that the
general solutions are subspaces of the Schwarzschild metric may be impulsory.
Indeed the Schwarzschild solution Eq.(\ref{eq:sch}), is a special case of the
general solutions Eq.(\ref{eq:alph}) for the case $\alpha =0$. We must notice
that both the Schwarzschild and the general solutions are in the same
Schwarzschild coordinate, which manifestly have different forms. Any
transformation to the new radial coordinate $r'=r+\alpha M$ requires the
Schwarzschild metric be written in this new coordinate too, which means we
should replace $r$ by $r'-\alpha M$. Thus in this new coordinate they will
have different forms too. As we know the Schwarzschild metric is an exterior
solution which is only valid for $r>2M$ and incomplete. This is true for the
general solutions but $r>(2-\alpha )M$ gives an extension larger than the
Schwarzschild solution and even with $\alpha >2$ the range of validity is
$r>0$. In the following we try to show how the spaces of general solutions
for $\alpha >2$ are complete.\\

A manifold endowed with an affine or metric geometry is said to be geodesically
complete if all geodesics emanating from any point can be extended to infinite
values of the affine parameters in both directions. In the case of a manifold
with a positive definite metric it can be shown \cite{kobo63} that geodesic
completeness and metric completeness are equivalent. Focusing on $\alpha >2$
which is the most likely values of it, the line element Eq.~(\ref{eq:alph})
is a vacuum solution abstracted away from any source for all values of $r$.
We also realize that
\[
g^{tt}=(1-{2M\over r+\alpha M})^{-1}\;
\]
is always positive for all $r>0$, so ``$t$'' is the invariant world time. The
hypersurfaces $t=const.$ are spacelike with a positive definite metric
\begin{equation}
d\sigma ^2=(1-{2M\over r+\alpha M})^{-1}dr^2
+(r+\alpha M)^2(d{\theta}^2+\sin^2\theta d\phi^2)\;
\end{equation}
which is a distance function. Since every Cauchy sequence with respect to this
distance function converges to a point in the manifold, it is metrically
complete.\\

Next we show that staticness is not a necessary initial condition if we merely
take $g_{\theta \theta}$ independent of time. Let's assume that the line
element has the following desired form
\begin{equation}
ds^2=B(r,t)dt^2-A(r,t)dr^2-D(r)({d\theta}^2+\sin^2\theta d\phi^2)\;
\label{eq:mti}
\end{equation}
Taking $ds^2=-g_{\mu \nu}dx^\mu dx^\nu $, the nonvanishing components of the
metric connection are
\begin{eqnarray}
\Gamma^t_{tt}&=&\frac{\dot{B}}{2B}\;,\;\;\;\;\;\;
\Gamma^t_{tr}=\frac{B'}{2B}\;,\;\;\;\;\;\;
\Gamma^t_{rr}=\frac{\dot{A}}{2B}\;,\nonumber\\
\Gamma^r_{tt}&=&\frac{B'}{2A}\;,\;\;\;\;\;\;
\Gamma^r_{tr}=\frac{\dot{A}}{2A}\;,\;\;\;\;\;\;
\Gamma^r_{rr}=\frac{A'}{2A}\;,\;\;\;\;\;\;
\Gamma^r_{\theta \theta}=-\frac{D'}{2A}\;,\;\;\;\;\;\;
\Gamma^r_{\phi \phi}=-\frac{D'}{2A}\sin^2\theta\;,\nonumber\\
\Gamma^\theta_{r\theta}&=&\frac{D'}{2D}\;,\;\;\;\;\;\;
\Gamma^\theta_{\phi \phi}=-\sin\theta \cos\theta\;,\nonumber\\
\Gamma^\phi_{r\phi}&=&\frac{D'}{2D}\;,\;\;\;\;\;\;
\Gamma^\phi_{\theta \phi}=\cot\theta.\;
\end{eqnarray}
where $(\;'\;)$ and $(\;\dot{} \;)$ denote derivatives with respect to $r$ and 
$t$ respectively. The nonvanishing components of the Ricci tensor are
\begin{eqnarray}
{\cal R}_{tt}&=&-\frac{B''}{2A}+\frac{B'}{4A}(\frac{A'}{A}+\frac{B'}{B})
-\frac{B'D'}{2AD}+\frac{\ddot{A}}{2A}-\frac{\dot{A}}{4A}
(\frac{\dot{A}}{A}+\frac{\dot{B}}{B})\\
{\cal R}_{tr}&=&-\frac{\dot{A}D'}{2AD}\\
{\cal R}_{rr}&=&\frac{B''}{2B}+\frac{D''}{D}-\frac{D'}{2D}
(\frac{D'}{D}+\frac{A'}{A})-\frac{B'}{4B}(\frac{A'}{A}+\frac{B'}{B})
-\frac{\ddot{A}}{2B}+\frac{\dot{A}}{4B}(\frac{\dot{A}}{A}+\frac{\dot{B}}{B})\\
{\cal R}_{\theta \theta}&=&-1+\frac{D'}{4A}(\frac{B'}{B}-\frac{A'}{A})
+\frac{D''}{2A}\\
{\cal R}_{\phi \phi}&=&\sin^2\theta {\cal R}_{\theta \theta}\nonumber\;
\end{eqnarray}

The Einstein field equations with $\Lambda=0$ for vacuum give
${\cal R}_{\mu \nu}=0$, so from setting ${\cal R}_{tr}=0$, we have
\begin{equation}
\dot{A}=0\;
\end{equation}
and with a reparametrization in the line element we may also have
\begin{equation}
\dot{B}=0\;
\end{equation}
This means that with the choice Eq.~(\ref{eq:mti}) as line element, the
solution is necessarily static.\\

By considering the bending of light, we try to find an upper bound for $\alpha $.
The deflection of light in a gravitational field predicted by the general
theory of relativity witnessed experimental verification in 1919. Since then,
there has been much studies on the {\it gravitational deflection of light by
the Sun} and {\it gravitational lensing} (GL). Under the great vision of
Zwicky \cite{zwic37}, observation of a QSO showed the first example of the GL
phenomena \cite{wals79}, and thereafter it has become the most important tool
for probing the universe. It is believed that GL can give valuable important
information on important questions, such as masses of galaxies and clusters of
galaxies, the existence of massive exotic objects, determination of cosmological
parameters and can be also used to test the alternative theories of gravitation
\cite{schn92}. The gravitational deflection of light has now been measured more
accurately at radio wavelenghts with using Very-Long-Baseline Interferometry
(VLBI), than at visible wavelenghts with available optical techniques.

Appealing to the spherically symmetric nature of the metric, we consider the
geodesics, without lose of generality, on the equatorial plane
$(\theta ={\pi \over2})$.
Following Weinberg \cite{weini72}, we get the equation for the photon
trajectories as
\begin{equation}
\label{eq:defeq}
\phi (r)-\phi_\infty = \int_r ^\infty {\biggl[
\frac{A(r)}{D(r)}\biggr] }^{1/2}
{\biggl[ \frac{D(r)}{D(r_o)}\frac{B(r_o)}{B(r)}-1\biggr] }^{-1/2}dr\;
\end{equation}
The Einstein deflection angle is given by
\begin{equation}
\Delta\phi =2|\phi (r_o)-\phi_\infty |-\pi\;
\end{equation}
In Eq.~(\ref{eq:alph}) we have
\begin{equation}
B=(1-\frac{2M}{r+\alpha M})\;,\;\;\;\;
A=(1-\frac{2M}{r+\alpha M})^{-1}\;,\;\;\;\;
D=(r+\alpha M)^2\; \label{eq:fal}
\end{equation}
which make the above integral well defined for $r_o>(3-\alpha )M$. Since the
radial coordinate is always non-negative, $\alpha $ merely take positive values.
Different values of $r_o$ and $\alpha $, yield different expansions for
$B$ and $D$, so we get different expressions for deflection angle. Our
investigation is on very small and sufficiently large values of $\alpha $,
since the rest of the interval brings no more to us. We give the details of
calculating in Appendix \ref{sec:deflec} and here use the results. For
$\alpha <1$, the Einstein deflection angle (up to the second order) is as
Eq.~(\ref{eq:defsm})
\begin{equation}
\label{eq:redef}
\Delta\phi =4x+4x^2\biggl[ \frac{15\pi }{16}-(1+\alpha )\biggr] +\cdots\;
\end{equation}
where
\[
x=\frac{MG}{r_o}\;
\]
and the resteriction imposed by the integral singularity is
\[
0<x<\frac{1}{3-\alpha }\;
\]
thus the closest approach is $r_o \approx (3-\alpha )MG$.
Switching off $\alpha $ in this solution one recovers the well-known
Schwarzschild solution, which has extensively examined by many authors (see
\cite{virell99} and refrences therein). However small values is qualitatively
similar (but quantitatively differnt) to the Schwarzschild case. Further
calculation shows that Eq.~(\ref{eq:redef}) is also valid for $\alpha >1$
(see Eq.~(\ref{eq:defla})), but it may not contain the closest approach.
Therefore, {\it for the weak field limit} (when $r_o$ is much larger than
$MG$) and {\it all possible values of $\alpha $}, we rewrite the equation as
\begin{equation}
\label{eq:defthe}
\Delta\phi =\Delta\phi_{fo}\biggl[ 1+\frac{MG}{r_o}\left(
\frac{15\pi }{16}-(1+\alpha )\right)\biggr] + \cdots\;
\end{equation}
where
\[
\Delta\phi_{fo}=\frac{4MG}{r_o}\;
\]
is the first order deflection angle.
The results of VLBI observations of extragalactic radio sources, show
radio-wave deflection by the Sun \cite{vlbi95} as
\begin{equation}
\label{eq:defobs}
\Delta\phi\approx\Delta\phi_{fo}(0.9998 \pm 0.0008)\;
\end{equation}
Since the order of magnitude of $MG/r_o$ for the Sun is $10^{-6}$,
Eq.~(\ref{eq:defthe}) and Eq.~(\ref{eq:defobs}) make an upper bound for
$\alpha $ as
\begin{equation}
\label{eq:defva}
0<\alpha <10^3\;
\end{equation}
For sufficiently large values of $\alpha $(actually$\alpha >3$) we may
obtain another expression for deflection angle (up to the second orther) in the
following form (see Eq.~(\ref{eq:defst}))
\begin{equation}
\label{eq:defcl}
\Delta \phi =8y+4y^2\biggl[ \frac{15\pi }{16}-5(\alpha -1)\biggr] +\cdots\;
\end{equation}
where
\[
y=\frac{r_o}{\alpha (\alpha -2)MG}\;
\]
and the range of validity is
\[
0<y<\frac{1}{\alpha }\;
\]
thus the closest approach is $r_o \approx 0$. Eq.~(\ref{eq:redef}) and
Eq.~(\ref{eq:defcl}) which contain the closest approach, can be used for
testing the general theory of relativity in a {\it strong gravitational field}.
Although, no test for the theory is known in this region, but there is an
open room for such investigations. Several possible observational candidates
have been proposed to test the Einstein theory of relativity in the vicinity
of a compact massive object. One of the current topics is the study of point
source lensing in the strong gravitational field region when the deflection
angle can be very large \cite{virb98}. Our calculation confirms that
deflection angle may take any small as well as large values depending on
$\alpha $ and it would provide a good key for the graviational lensing
studies.\\
Consequently bending of light phenomena, both in the {\it weak field} and
{\it strong field} limits, restrict all non-negative values of $\alpha $ as 
Eq.~(\ref{eq:defva}) up to $10^3$.\\

Now we would like to use observational data of the {\it precession
of perihelia} measurements, in order to find a better bound for $\alpha $.
Following Weinberg \cite{weinii72}, for a test particle moves on a timelike
geodesic in the $\theta =\pi /2$ plane, the angle swept is given by
\begin{equation}
\label{eq:preeq}
\phi(r)-\phi(r_-)=\int_{r_-}^r dr \biggl[
\frac{A^{1/2}(r)}{D(r)}\biggr]
\biggl[
{D_-(B^{-1}(r)-B_-^{-1})-D_+(B^{-1}(r)-B_+^{-1}) \over
D_+D_-(B_+^{-1}-B_-^{-1})}-{1\over D(r)}\biggr]^{-1/2}\nonumber\\
\end{equation}
where
\[
D_{\pm}=D(r_{\pm}) \;\;\;\;\; \& \;\;\;\;\; B_{\pm}=B(r_{\pm})\;
\]
The orbit precesses in each revolution by an angle
\begin{equation}
\Delta \phi =2| \phi (r_+) - \phi (r_-)| -2\pi\;
\end{equation}
By using the metric components presented in Eq.~(\ref{eq:fal}), we have gotten
the expression for the precession per revolution (up to the second order)
which the details are given in Appendix \ref{sec:preces}. So we may write
Eq.~(\ref{eq:prec}) in here as
\begin{equation}
\Delta \phi = \Delta \phi_{fo} \biggl[ 1+\frac{MG}{L} \biggl( \frac{19}{6}+
\frac{{\it e}^2}{4}-2\alpha (1+{\it e}^2) \biggr)
\biggr] + \cdots\; \label{eq:preth}
\end{equation}
where $``L''$ and $``{\it e}''$ are the semilatus rectum and eccentricity
respectively, and $ \Delta \phi_{fo}$ is the well-known first order
approximation
\[
\Delta \phi_{fo} =\frac{6\pi MG}{L}\;
\]
Fortunately by development of Long-Baseline radio Interferometry and analysis
of Radar Ranging Data, there are accurate measurements of precession which
typically show \cite{alto76}
\begin{equation}
\Delta \phi \approx \Delta \phi_{fo} (1.003 \pm 0.005)\;
\end{equation}
Thus matching the theory with observation, with using typical values of
$\frac{M G}{L}$ and {\it e}, we get an upper bound for $\alpha $ as
\begin{equation}
\label{eq:acuva}
0<\alpha <10^5\;
\end{equation}

We conclude that measurements of these two tests of the general theorey of
relativity in the {\it weak field limit} according to Eq.~(\ref{eq:defva}) and
Eq.~(\ref{eq:acuva}), restrict the allowed values of $\alpha $ to $10^3$.
Though the observation of the GL phenomena is a difficult task, it would
support our presented metric components role in a {\it strong gravitational
field} and would also give an accurate bound for $\alpha $.\\

By calculating the Riemann tensor scalar invariant, we receive useful
information about the existence of singularities. For the line element
Eq.~(\ref{eq:alph}) in which $\alpha \neq0$, it is equal to
\begin{equation}
{\cal R}^a\;_{bcd}{\cal R}_a\;^{bcd}=\frac{48M^2}{D^3}=\frac{48M^2}
{(r+\alpha M)^6}\;
\label{eq:mcur}
\end{equation}
As it is evident, the presence of $\alpha $ makes the scalar finite in the
whole range of $r$, meaning that the solutions are free of any intrinsic
singularity. Meanwhile there may be a coordinate singularity at
$r=(2-\alpha )M$ according to the restriction imposed by
Eq.~(\ref{eq:defva}).\\

We would like to mention two points about this work. One is that, sometimes in
literatures for discussing this problem, coordinate $r$ is defined so that the
area of the surfaces $r= const.$ to be $4 \pi r^2$ \cite{rindI77}. This
generally is not the case, since before fixing the metric there is no
possibility of speaking the distance, then in the same way, there is no
possibility of speaking the area. We take $r=(x^2+y^2+z^2)^{1/2}$ where
$(x, y, z)$ are usual cartesian space coordinates.

The other point which should be taken with caution is, while at first $r$ was
taken as a space coordinate with the range $(0,\infty )$ and the particle was
located at $r=0$, at end we come to the conclusion that r is merely a space
coordinate in the interval $((2-\alpha )M,\infty )$. For the rest of the
interval $(0,(2-\alpha )M)$, it is standing as a time coordinate. This
contradiction or at least ambiguity raises the question that {\it While the
location of the particle is not well-defined, how may we speak of the value
of $D$ at this point?} This ambiguity should be found a satisfactory
explanation.\\

\section{\bf Case\ ${\bf \Lambda \neq0}$ , ${\bf \alpha =0}$}

Recently for vaccum sphericaly symmetric space, a non-static solution of
Einstein field equations with cosmological constant in a proper FRW type
coordinate system has been cosidered \cite{abbas99}. The result shows a
singularity at the origin which the intrinsic nature of it may be checked by
calculating the invariant Gaussian curvature or the Riemann tensor scalar
invariant. In this section we choose tensor analysis to find the answer.
So we start with the following form of the line element
\begin{equation}
ds^2=B(r,t)dt^2-R^2(t)\biggl[A(r,t)dr^2+r^2
({d\theta}^2+\sin^2\theta d\phi^2)\biggr]\;
\label{eq:nsss}
\end{equation}
where $R(t)$ is the scale factor and satisfies the Friedmann equation
\begin{equation}
\frac{\dot {R^2}}{R^2}=\frac{\ddot {R}}{R}=\frac{\Lambda}{3}\;
\label{eq:frid}
\end{equation}
and $\Lambda $ is the cosmological constant.
The cosmological time $t$, which determines the evolution of the universe is
measured by the proper time of clocks fixed on the geodesically moving galaxies
comprising the universe, synchronized such that the time $t=0$ corresponds to
the beginning of the universe at the big bang.

According to Ref.~\cite{abbas99} the nonvanishing components of the metric
connection are
\begin{eqnarray}
\Gamma^t_{tt}&=&\frac{\dot{B}}{2B}\;,\;\;\;\;\;
\Gamma^t_{tr}=\frac{B'}{2B}\;,\;
\Gamma^t_{rr}=\frac{R\dot{R} A}{B}+\frac{R^2 \dot{A}}{2B}\;,\;
\Gamma^t_{\theta \theta}=\frac{R\dot{R} r^2}{B}\;,\;
\Gamma^t_{\phi \phi}=\frac{R\dot{R} r^2}{B}\sin^2\theta\;,\nonumber\\
\Gamma^r_{tt}&=&\frac{B'}{2AR^2}\;,\;
\Gamma^r_{tr}=\frac{\dot{R}}{R}+\frac{\dot{A}}{2A}\;,\;\;\;\;
\Gamma^r_{rr}=\frac{A'}{2A}\;,\;\;\;\;\;\;
\Gamma^r_{\theta \theta}=-\frac{r}{A}\;,\;\;\;\;
\Gamma^r_{\phi \phi}=-\frac{r}{A}\sin^2\theta\;,\nonumber\\
\Gamma^\theta_{t\theta}&=&\frac{\dot{R}}{R}\;,\;\;\;\;\;\;\;
\Gamma^\theta_{r\theta}=\frac{1}{r}\;,\;\;\;\;\;\;\;\;\;\;\;\;\;\;
\Gamma^\theta_{\phi \phi}=-\sin\theta \cos\theta\;,\nonumber\\
\Gamma^\phi_{t\phi}&=&\frac{\dot{R}}{R}\;,\;\;\;\;\;\;\;
\Gamma^\phi_{r\phi}=\frac{1}{r}\;,\;\;\;\;\;\;\;\;\;\;\;\;\;\;
\Gamma^\phi_{\theta \phi}=\cot\theta.\;\label{eq:gnsss}
\end{eqnarray}
The nonvanishing components of the curvature tensor then become
\begin{eqnarray}
{\cal R}^t\;_{rtr}&=&\frac{A\dot{R^2}}{B}+\frac{\ddot{A}R^2}{2B}-\frac{B''}{2B}
+\frac{3\dot{A}R\dot{R}}{2B}\;,\;\;\;\;\;\;\;
{\cal R}^r\;_{trt}=\frac{B''}{2AR^2}-\frac{\ddot{A}}{2A}
-\frac{3\dot{A}\dot{R}}{2AR}-\frac{(\dot{R})^2}{R^2}\;,\nonumber\\
{\cal R}^t\;_{\theta t\theta}&=&\frac{A'r}{2A^2}+\frac{r^2R\dot{R}}{B}
(\frac{\dot{A}}{2A}+\frac{\dot{R}}{R})\;,\;\;\;\;\;\;\;\;\;\;\;\;\;\;\;
{\cal R}^r\;_{\theta r\theta}={\cal R}^t\;_{\theta t\theta}\;,\nonumber\\
{\cal R}^t\;_{\phi t\phi}&=&{\cal R}^t\;_{\theta t\theta}\sin^2\theta\;,
\;\;\;\;\;\;\;\;\;\;\;\;\;\;\;\;\;\;\;\;\;\;\;\;\;\;\;\;\;\;\;\;\;
{\cal R}^r\;_{\phi r\phi}={\cal R}^t\;_{\phi t\phi}\;,\nonumber\\
\nonumber\\
{\cal R}^\theta\;_{t\theta t}&=&\frac{B'}{2rAR^2}+\frac{\dot{R}}{R}
(\frac{\dot{B}}{2B}-\frac{\dot{R}}{R})\;,\;\;\;\;\;\;\;\;\;\;\;\;\;\;\;\;
{\cal R}^\phi\;_{t\phi t}={\cal R}^\theta\;_{t\theta t}\;,\nonumber\\
{\cal R}^\theta\;_{r\theta r}&=&\frac{A'}{2rA}+\frac{\dot{R}}{R}
(\frac{AR\dot{R}}{B}+\frac{R^2\dot{A}}{2B})\;,\;\;\;\;\;\;\;\;\;\;\;\;
{\cal R}^\phi\;_{r\phi r}={\cal R}^\theta\;_{r\theta r}\;,\nonumber\\
{\cal R}^\theta\;_{\phi \theta \phi}&=&{\cal R}^\phi\;_{\theta \phi \theta}\sin^2\theta\;,
\;\;\;\;\;\;\;\;\;\;\;\;\;\;\;\;\;\;\;\;\;\;\;\;\;\;\;\;\;\;\;
{\cal R}^\phi\;_{\theta \phi \theta}=1-\frac{1}{A}+\frac{r^2\dot{R^2}}{B}\;.\;
\label{eq:cute}
\end{eqnarray}
Since the metric components have no cross terms and furthermore the curvature
tensors have some algebraic properties, the Riemann tensor scalar invariant
takes a simpler form as
\begin{equation}
{\cal R}^a\;_{bcd}{\cal R}_a\;^{bcd}=
2\sum_{\mu \neq \nu}(g^{\mu \mu}{\cal R}^\nu\;_{\mu \nu \mu})^2\;
\end{equation}
Using Eq.~(\ref{eq:cute}) gives
\begin{eqnarray}
{\cal R}^a\;_{bcd}{\cal R}_a\;^{bcd}&=&
4\biggl(\frac{B''}{2ABR^2}-\frac{\ddot{A}}{2AB}-\frac{3\dot{A}\dot{R}}{2ABR}
-\frac{\dot{R^2}}{BR^2}\biggr)^2\nonumber\\
&&+ 16\biggl(\frac{A'}{2rA^2R^2}+\frac{\dot{A} \dot{R}}{2ABR}
+\frac{\dot{R^2}}{BR^2}\biggr)^2\nonumber\\
&&+ 4\biggl(\frac{1}{r^2R^2}-\frac{1}{r^2AR^2}+\frac{\dot{R^2}}{BR^2}\biggr)^2\;
\label{eq:sccur}
\end{eqnarray}
The $rr$ component of the field equation gives
\[
{\cal R}_{rr}+\Lambda g_{rr}=0\;
\]
and makes the equall terms for the first parentheses as
\begin{equation}
\frac{B''}{2ABR^2}-\frac{\ddot{A}}{2AB}-\frac{3\dot{A}\dot{R}}{2ABR}
-\frac{\dot{R^2}}{BR^2}=-\biggl(\frac{1}{r^2R^2}-\frac{1}{r^2AR^2}
+\frac{\dot{R^2}}{BR^2}\biggr)\;
\label{eq:equfit}
\end{equation}
The $\theta \theta$ component gives
\[
{\cal R}_{\theta \theta}+\Lambda g_{\theta \theta}=0\;
\]
which this leads to
\begin{equation}
\frac{A'}{2rA^2R^2}+\frac{\dot{A} \dot{R}}{2ABR}+\frac{\dot{R^2}}{BR^2}=
-\frac{1}{2}\biggl(\frac{1}{r^2R^2}-\frac{1}{r^2AR^2}
+\frac{\dot{R^2}}{BR^2}-\frac{3\dot{R^2}}{R^2}\biggr)\;
\label{eq:eqused}
\end{equation}
By solving field equations we also have \cite{abbas99}
\begin{equation}
1-\frac{1}{A}+\frac{\Lambda}{3}r^2R^2(A-1)=\frac{2M}{rR}\;\label{eq:nonab1}
\end{equation}
or
\begin{equation}
B=A^{-1}=\frac{1}{2}\biggl[{\sqrt{(1-\frac{2M}{rR}
-\frac{\Lambda}{3}r^2R^2)^2+\frac{4\Lambda}{3}r^2R^2}}
+(1-\frac{2M}{rR}-\frac{\Lambda}{3}r^2R^2)\biggr]\;\label{eq:nonab2}
\end{equation}
Inserting ~(\ref{eq:equfit}),~(\ref{eq:eqused}) and ~(\ref{eq:nonab1}) in
~(\ref{eq:sccur}), gives
\begin{equation}
{\cal R}^a\;_{bcd}{\cal R}_a\;^{bcd}=\frac{48M^2}{r^6 R^6(t)}
+\frac{24}{9}\Lambda^2\;
\label{eq:nscur}
\end{equation}
This evidently exhibits the existence of an intrinsic singularity at the
origin. Removing this deficiency, leads us to the most general form of the
solutions which comes next. As it is expected, Eq.~(\ref{eq:nscur}) with
$R(t)=1$ gives the result of the static case \cite{larod77}.\\

We end this section with emphasizing on some aspects of Eq.~(\ref{eq:nonab2})
and Eq.~(\ref{eq:nscur}).
Firstly, the existence of a nonzero cosmological constant regardless
of its actual value, is sufficient to prevent from Schwarzschild singularity
at $r \approx 2M$. Recent observations of type Ia supernovae indicating a
universal expansion, put forward the possible existence of a small positive
cosmological constant \cite{pococo98}. These evidences persuade us that in a
$\Lambda $-dominated universe, we would have no troubles in describing the
whole space.

Secondly, since there is no singularity for $r>0$, then there is no ambiguity
in defining coordinate $r$, which mentioned at the end of Sec.II. {\it ``$r$''
is a space coordinate in the whole interval $(0, \infty )$ and we may speak of
the value of D at $r=0$ without any problem}.

Thirdly, a coordinate transformation \cite{abbas99}, transforms the above
non-static metric to the well-known {\it Schwarzschild-de Sitter} static metric
Eq.~(\ref{eq:scds}).
This metric has some deficiencies which we would like to discuss briefly.
Despite of an intrinsic singularity at the origin which is similar to our case,
it has a coordinate type singularity at $r \approx \sqrt {\frac{3}{\Lambda}}$
in a $\Lambda $-dominated universe. Though the presence of cosmological
constsnt removes the Schwarzschild singularity from our metric, but the
transformed form of it, {\it i.e.} Schwarzschild-de Sitter metric, has again
the problem of exchanging the meaning of space and time. On the other hand when
$M=0$, the assumed FRW background could not be revisited but the attached
homogenity and isotropy would resist. And more importantly, this metric shows
a redshift-magnitude relation which contradicts the observational data
\cite{abbkh98}.

Therefore it is adequate to discard the Schwarzschild-de Sitter metric, in
favor of our presented metric, as a proper frame of reference in the presence
of $\Lambda $, since all of the mentioned deficiencies do not exist in it.
The metric asymptotically approaches to the non-static de Sitter metric
which is appropriate for a $\Lambda $-dominated universe.
Furthermore as we show next, the presented metric has the suitability even to
remove the intrinsic singularity at the origin.\\

\section{\bf Case\ ${\bf \Lambda \neq0}$ , ${\bf \alpha \neq0}$}

Since it turns out that there is an intrinsic singularity with the choice
$\alpha =0$, we would like to solve the problem by using the most general form
of the line element. In this section we work out expressions for metric
coeficients in the FRW universe and obtain an analytic metric everywhere.
Therefore we choose the metric for the universe in terms of the coordinates
$(r,t)$ to be
\begin{equation}
ds^2=B(r,t)dt^2-R^2(t)\biggl[A(r,t)dr^2+D(r)
({d\theta}^2+\sin^2\theta d\phi^2)\biggr]\;
\label{eq:frw}
\end{equation}
The nonvanishing components of the metric and metric connection respectively
are
\begin{equation}
g_{tt}=-B,\;\;\;\;\;\;\;\; g_{rr}=AR^2,\;\;\;\;\;\;\;\;
g_{\theta \theta}=DR^2,\;\;\;\;\;\;\;\; g_{\phi \phi}=DR^2\sin^2\theta\;
\end{equation}
\begin{eqnarray}
\Gamma^t_{tt}&=&\frac{\dot{B}}{2B}\;,\;\;\;\;\;
\Gamma^t_{tr}=\frac{B'}{2B}\;,\;
\Gamma^t_{rr}=\frac{R\dot{R} A}{B}+\frac{R^2 \dot{A}}{2B}\;,\;
\Gamma^t_{\theta \theta}=\frac{R\dot{R} D}{B}\;,\;
\Gamma^t_{\phi \phi}=\frac{R\dot{R} D}{B}\sin^2\theta\;,\nonumber\\
\Gamma^r_{tt}&=&\frac{B'}{2AR^2}\;,\;
\Gamma^r_{tr}=\frac{\dot{R}}{R}+\frac{\dot{A}}{2A}\;,\;\;\;\;
\Gamma^r_{rr}=\frac{A'}{2A}\;,\;\;\;\;\;\;
\Gamma^r_{\theta \theta}=-\frac{D'}{2A}\;,\;\;
\Gamma^r_{\phi \phi}=-\frac{D'}{2A}\sin^2\theta\;,\nonumber\\
\Gamma^\theta_{t\theta}&=&\frac{\dot{R}}{R}\;,\;\;\;\;\;\;\;
\Gamma^\theta_{r\theta}=\frac{D'}{2D}\;,\;\;\;\;\;\;\;\;\;\;\;
\Gamma^\theta_{\phi \phi}=-\sin\theta \cos\theta\;,\nonumber\\
\Gamma^\phi_{t\phi}&=&\frac{\dot{R}}{R}\;,\;\;\;\;\;\;\;
\Gamma^\phi_{r\phi}=\frac{D'}{2D}\;,\;\;\;\;\;\;\;\;\;\;\;
\Gamma^\phi_{\theta \phi}=\cot\theta.\;\label{eq:gfrw}
\end{eqnarray}
where $(\;'\;)$ and $(\;\dot{} \;)$ denote derivatives with respect to $r$ and
$t$ respectively. The nonvanishing components of the Ricci tensor then become
\begin{eqnarray}
{\cal R}_{tt}&=&\frac{B}{AR^2}\biggl[-\frac{B''}{2B}+\frac{B'}{4B}
(\frac{A'}{A}+\frac{B'}{B})-\frac{B'D'}{2BD}\biggr]+\frac{\ddot{A}}{2A}
-\frac{\dot{A}}{4A}(\frac{\dot{A}}{A}+\frac{\dot{B}}{B})\nonumber\\
&&+ \frac{3\ddot{R}}{R}-\frac{\dot{R}}{R}(\frac{3\dot{B}}{2B}
-\frac{\dot{A}}{A})\\
{\cal R}_{tr}&=&-\frac{\dot{A}D'}{2AD}-\frac{B'\dot{R}}{BR}\\
{\cal R}_{rr}&=&\frac{B''}{2B}+\frac{D''}{D}-\frac{D'}{2D}
(\frac{D'}{D}+\frac{A'}{A})-\frac{B'}{4B}(\frac{A'}{A}+\frac{B'}{B})\nonumber\\
&&- \frac{R^2A}{B}\biggl[\frac{\ddot{R}}{R}+\frac{\dot{R}}{R}
(\frac{2\dot{R}}{R}+\frac{2\dot{A}}{A}-\frac{\dot{B}}{2B})
+\frac{\ddot{A}}{2A}-\frac{\dot{A}}{4A}
(\frac{\dot{A}}{A}+\frac{\dot{B}}{B})\biggr]\\
{\cal R}_{\theta \theta}&=&-1+\frac{D'}{4A}(\frac{B'}{B}-\frac{A'}{A})
+\frac{D''}{2A}
-\frac{R^2D}{B}\biggl[\frac{\ddot{R}}{R}+\frac{\dot{R}}{R}
(\frac{2\dot{R}}{R}+\frac{\dot{A}}{2A}-\frac{\dot{B}}{2B})\biggr]\\
{\cal R}_{\phi \phi}&=&\sin^2\theta {\cal R}_{\theta \theta}\nonumber\;
\end{eqnarray}

To solve the vacuum field equations ${\cal R}_{\mu \nu}+\Lambda g_{\mu \nu}=0$,
we first begin with ${\cal R}_{tr}=0$ and introduce a new variable $\rho$ to be
\begin{equation}
\rho = R(t)D^{1/2}(r)\;
\end{equation}
From this equation we obtain
\begin{equation}
\frac{1}{2}\dot{R}D'D^{-1/2}(\frac{A^*}{A}+\frac{B^*}{B})=0\;
\end{equation}
which $( ^* )$ means differentiation with respect to $\rho$. Since $\dot{R}$
and $D'\neq0$, we should have
\begin{equation}
\frac{A^*}{A}+\frac{B^*}{B}=0\;
\end{equation}
Integration with respect to $\rho$ and imposing flat boundary condition
at large distances yields
\begin{equation}
AB=1\;
\end{equation}
in agreement with the FRW background.
In the next step, let's consider
\begin{equation}
\frac{{\cal R}_{tt}}{B}+\frac{{\cal R}_{rr}}{AR^2}=0\;
\end{equation}
which we get
\[
\frac{-D'^2}{2AD^2 R^2}+\frac{D''}{ADR^2}=0\;
\]
or we may write
\begin{equation}
(D'D^{-1/2})'=0\;
\end{equation}
Integration with respect to $r$ gives
\begin{equation}
D^{1/2}=r+\alpha M \;\;\;\;\;\mbox{or}\;\;\;\;\; D=(r+\alpha M)^2\;
\end{equation}
As before $\alpha$ is a positive constant in the range
$0<\alpha <10^3$.

Finally from the $\theta \theta$ component of the field equation,
${\cal R}_{\theta \theta}+\Lambda g_{\theta \theta}=0$, we obtain the
functional form of $A(r,t)$ as follows
\[
-1+\frac{1}{A}-\rho \frac{A^*}{A^2}-\Lambda \rho^2(A-1)-\frac{\Lambda}{3}
\rho^3 A^*=0\;
\]
or
\begin{equation}
\frac{d}{d\rho}\biggl[\rho(1-\frac{1}{A})\biggr]
+\frac{\Lambda}{3}\frac{d}{d\rho}
\biggl[\rho^3 (A-1)\biggr]=0\;
\end{equation}
Integration then yields
\begin{equation}
\rho(1-\frac{1}{A})+\frac{\Lambda}{3}\rho^3 (A-1)=c\;\label{eq:fufa}
\end{equation}
where $c$ is constant of integration. Comparing our result with post-Newtonian
limit gives $c=2M$. Finally the result is
\[
A=B^{-1}=(\frac{2\Lambda}{3}\rho^2)^{-1}\biggl[{\sqrt{(1-\frac{2M}{\rho}
-\frac{\Lambda}{3}\rho^2)^2+\frac{4\Lambda}{3}\rho^2}}
-(1-\frac{2M}{\rho}-\frac{\Lambda}{3}\rho^2)\biggr]\;
\]
or
\begin{equation}
B=A^{-1}=\frac{1}{2}\biggl[{\sqrt{(1-\frac{2M}{\rho}
-\frac{\Lambda}{3}\rho^2)^2+\frac{4\Lambda}{3}\rho^2}}
+(1-\frac{2M}{\rho}-\frac{\Lambda}{3}\rho^2)\biggr]\;
\end{equation}

As it is evident from the functional form of A, B and D, this metric has no
apparent singularity and a straightforward calculation gives the Riemann
tensor scalar invariant in the form
\begin{equation}
{\cal R}^a\;_{bcd}{\cal R}_a\;^{bcd}=\frac{48M^2}{D^3 R^6(t)}
+\frac{24}{9}\Lambda^2\;
\label{eq:gnscur}
\end{equation}
If we set $\Lambda=0, R(t)=1$ in ~(\ref{eq:gnscur}), Eq.~(\ref{eq:mcur}) will
be obtained, and furthermore with $D=r^2$, Eq.~(\ref{eq:cur}) would regain.\\

It is remarkable that having $\alpha \neq0$, we would have a well-defined
metric in the whole space which asymptotically approaches to the non-static
de Sitter metric; {\it i.e.} the appropriate metric for a $\Lambda $-dominated
universe.\\

\section{\bf Geodesic Equations}

Our next task is to obtain and solve the geodesic equations of a freely falling
material particle in a proper FRW type coordinate system. In the static case,
the basic equations which determine the geodesic structure of the manifold have
been fully developed \cite{larod77,hlrd74}. We solve the problem in the
non-static case by using the equations of free fall
\begin{equation}
\frac{d^2x^\mu}{ds^2}+\Gamma^\mu_{\nu \lambda}\frac{dx^\nu}{ds}\frac{dx^\nu}{ds}=0\;
\end{equation}
Using the nonvanishing components of affine connection, given by
Eq.~(\ref{eq:gfrw}) we have
\begin{eqnarray}
\frac{d^2t}{ds^2}&+&\frac{\dot{B}}{2B}(\frac{dt}{ds})^2
+\frac{B'}{B}\frac{dt}{ds}\frac{dr}{ds}+(\frac{R\dot{R}A}{B}
+\frac{R^2\dot{A}}{2B})(\frac{dr}{ds})^2
+\frac{R\dot{R}D}{B}(\frac{d\theta}{ds})^2\nonumber\\
&+& \frac{R\dot{R}D}{B}\sin^2\theta(\frac{d\phi}{ds})^2=0\\
\frac{d^2r}{ds^2}&+&\frac{B'}{2AR^2}(\frac{dt}{ds})^2
+2(\frac{\dot{R}}{R}+\frac{\dot{A}}{2A})\frac{dt}{ds}\frac{dr}{ds}
+\frac{A'}{2A}(\frac{dr}{ds})^2-\frac{D'}{2A}(\frac{d\theta}{ds})^2\nonumber\\
&-& \frac{D'}{2A}\sin^2\theta(\frac{d\phi}{ds})^2=0\\
\label{eq:geth}
\frac{d^2\theta}{ds^2}&+&2\frac{\dot{R}}{R}\frac{dt}{ds}\frac{d\theta}{ds}
+\frac{D'}{D}\frac{dr}{ds}\frac{d\theta}{ds}-\sin\theta \cos\theta
(\frac{d\phi}{ds})^2=0\\
\frac{d^2\phi}{ds^2}&+&2\frac{\dot{R}}{R}\frac{dt}{ds}\frac{d\phi}{ds}
+\frac{D'}{D}\frac{dr}{ds}\frac{d\phi}{ds}+2\cot\theta \frac{d\theta}{ds}
\frac{d\phi}{ds}=0\;
\end{eqnarray}

Since the field is isotropic, we may consider the orbit of our particle to be
confined to the equatorial plane, that is $\theta={\pi \over2}$. Then
Eq.~(\ref{eq:geth}) immediately is satisfied and we may forget about $\theta$
as a dynamical variable. Introducing the previous variable ``$\rho$'' and
using the Friedmann equation, make the above equations as
\begin{eqnarray}
\label{eq:geti}
\frac{d^2t}{ds^2}&+&\sqrt{\frac{\Lambda}{3}} \rho
(\frac{A^*}{2A}+\frac{\Lambda}{3} \rho A^2+\frac{\Lambda}{6} \rho^2 AA^*)
(\frac{dt}{ds})^2-(\frac{A^*}{A}+2\frac{\Lambda}{3} \rho A^2
+\frac{\Lambda}{3} \rho^2 AA^*)\frac{dt}{ds}\frac{d\rho}{ds}\nonumber\\
&+&\sqrt{\frac{\Lambda}{3}}(A^2+\frac{1}{2} \rho AA^*)(\frac{d\rho}{ds})^2
+\sqrt{\frac{\Lambda}{3}} \rho^2 A(\frac{d\phi}{ds})^2=0\\
\label{eq:gero}
\frac{d^2\rho}{ds^2}&-&\sqrt{\frac{\Lambda}{3}} \rho \frac{d^2t}{ds^2}
-(\frac{\Lambda}{3} \rho +\frac{A^*}{2A^3}+\frac{\Lambda}{6} \rho^2 \frac{A^*}{A})
(\frac{dt}{ds})^2
+\frac{A^*}{2A}(\frac{d\rho}{ds})^2-\frac{\rho}{A}(\frac{d\phi}{ds})^2=0\\
\frac{d^2\phi}{ds^2}&+&\frac{2}{\rho}\frac{d\rho}{ds}\frac{d\phi}{ds}=0\;
\end{eqnarray}
The last equation gives
\begin{equation}
\frac{d}{ds}(\rho^2 \frac{d\phi}{ds})=0\;
\label{eq:dds}
\end{equation}
Integrating ~(\ref{eq:dds}) with respect to $s$ we get
\begin{equation}
\frac{d\phi}{ds}=\frac{J}{\rho^2}\;
\end{equation}
where $J$ is constant of integration.\\

In order to derive the exact solution, first rewrite Eq.~(\ref{eq:geti}) as
\begin{eqnarray}
\label{eq:mgeti}
\frac{d}{ds}\biggl[(\frac{1}{A}-\frac{\Lambda}{3}\rho^2 A)\frac{dt}{ds}\biggr]
&+&\sqrt{\frac{\Lambda}{3}} \rho A\biggl[
\sqrt{\frac{\Lambda}{3}} \rho \frac{d^2t}{ds^2}
+(\frac{A^*}{2A^3}+\frac{\Lambda}{3} \rho 
+\frac{\Lambda}{6} \rho^2 \frac{A^*}{A})(\frac{dt}{ds})^2
+\frac{J^2}{A\rho^3}\biggr]\nonumber\\
&+&\sqrt{\frac{\Lambda}{3}}(A+\frac{\rho}{2}A^*)(\frac{d\rho}{ds})^2=0\;
\end{eqnarray}
Substituting Eq.~(\ref{eq:gero}) into Eq.~(\ref{eq:mgeti}) yields
\[
\frac{d}{ds}\biggl[(\frac{1}{A}-\frac{\Lambda}{3}\rho^2 A)\frac{dt}{ds}\biggr]
+\sqrt{\frac{\Lambda}{3}}\biggl[\rho A\frac{d^2\rho}{ds^2}
+(A+\rho A^*)(\frac{d\rho}{ds})^2\biggr]=0\;
\]
or
\begin{equation}
\frac{d}{ds}\biggl[(\frac{1}{A}-\frac{\Lambda}{3}\rho^2 A)\frac{dt}{ds}
+\sqrt{\frac{\Lambda}{3}}\rho A\frac{d\rho}{ds}\biggr]=0\;
\end{equation}
which can be integrated to get
\begin{equation}
(\frac{1}{A}-\frac{\Lambda}{3}\rho^2 A)\frac{dt}{ds}
+\sqrt{\frac{\Lambda}{3}}\rho A\frac{d\rho}{ds}=c_1\;\label{eq:sgeq}
\end{equation}
where $c_1$ is constant of integration. Derivative of $\rho$ with respect to
$s$ gives
\begin{equation}
\frac{d\rho}{ds}=\sqrt{\frac{\Lambda}{3}}\rho \frac{dt}{ds}+R\frac{dr}{ds}\;
\label{eq:adds}
\end{equation}
if $r\to \infty$ or $\rho \to \infty$, then ~(\ref{eq:adds}) reduces to
\[
\frac{d\rho}{ds}\to \sqrt{\frac{\Lambda}{3}}\rho\;
\]
because for free fall at infinity we have
\[
\frac{dr}{ds}=0\;\;\;\;\; \& \;\;\;\;\; \frac{dt}{ds}=1\;
\]
Therefore $c_1$ would be fixed by asymptotic behavior as $c_1=1$.
Eq.~(\ref{eq:sgeq}) then becomes
\begin{equation}
\frac{dt}{ds}=\frac{1-\sqrt{\frac{\Lambda}{3}}\rho A\frac{d\rho}{ds}}
{{1\over A}-\frac{\Lambda}{3}\rho^2 A}\;
\end{equation}

We have yet another equation, the line element equation, which in terms
of $\rho$ variable has the form
\begin{equation}
\label{eq:leeq}
(\frac{1}{A}-\frac{\Lambda}{3} \rho^2 A)(\frac{dt}{ds})^2
+2\sqrt{\frac{\Lambda}{3}}\rho A \frac{dt}{ds}\frac{d\rho}{ds}
-A(\frac{d\rho}{ds})^2-\rho^2 (\frac{d\phi}{ds})^2=1\;
\end{equation}
Rewriting Eq.~(\ref{eq:geti}) as
\begin{eqnarray}
\frac{d^2t}{ds^2}&+&\sqrt{\frac{\Lambda}{3}}\rho\frac{A^*}{2A}(\frac{dt}{ds})^2
-\frac{A^*}{A}\frac{dt}{ds}\frac{d\rho}{ds}\nonumber\\
&+&\sqrt{\frac{\Lambda}{3}} A\biggl[\frac{\Lambda}{3} \rho^2 A(\frac{dt}{ds})^2
-2\sqrt{\frac{\Lambda}{3}} \rho A\frac{dt}{ds}\frac{d\rho}{ds}
+A(\frac{d\rho}{ds})^2+\frac{J^2}{\rho^2}\biggr]\nonumber\\
&+& \frac{1}{2}\sqrt{\frac{\Lambda}{3}}\rho A^*\biggl[
\frac{\Lambda}{3} \rho^2 A(\frac{dt}{ds})^2
-2\sqrt{\frac{\Lambda}{3}} \rho A\frac{dt}{ds}\frac{d\rho}{ds}
+A(\frac{d\rho}{ds})^2\biggr]=0\;
\end{eqnarray}
and using Eq.~(\ref{eq:leeq}), we come to
\begin{equation}
\frac{d^2t}{ds^2}+\sqrt{\frac{\Lambda}{3}}
(1+\rho\frac{A^*}{A})(\frac{dt}{ds})^2
-\frac{A^*}{A}\frac{dt}{ds}\frac{d\rho}{ds}
-\sqrt{\frac{\Lambda}{3}}(A+\frac{\rho}{2}A^*+\frac{J^2}{2\rho} A^*)=0\;
\end{equation}
multiplying the above equation by $\sqrt{\frac{\Lambda}{3}}\rho$ and inserting
the result in Eq.~(\ref{eq:gero}) we find
\begin{eqnarray}
\frac{d^2\rho}{ds^2}&-&\frac{A^*}{2A^2}\biggl[
(\frac{1}{A}-\frac{\Lambda}{3}\rho^2 A)(\frac{dt}{ds})^2
+2\sqrt{\frac{\Lambda}{3}}\rho A\frac{dt}{ds}\frac{d\rho}{ds}
-A(\frac{d\rho}{ds})^2\biggr]-\frac{J^2}{\rho^3 A}\nonumber\\
&-& \frac{\Lambda}{3}\rho (A+\frac{\rho}{2}A^*
+\frac{J^2}{2\rho} A^*)=0\;
\end{eqnarray}
From Eq.~(\ref{eq:leeq}), we notice that the expression in the bracket is just
equal to $1+\frac{\ J^2}{\ \rho^2}$, so we obtain a second order differential
equation for $\rho$ as
\begin{equation}
\frac{d^2\rho}{ds^2}-\frac{A^*}{2A^2}(1+\frac{J^2}{\rho^2})
-\frac{J^2}{\rho^3 A}
-\frac{\Lambda}{3}\rho (A+\frac{\rho}{2}A^*+\frac{J^2}{2\rho} A^*)=0\;
\end{equation}
which can be written
\begin{equation}
\frac{d^2\rho}{ds^2}+\frac{1}{2}\frac{d}{d\rho}\biggl[
(\frac{1}{A}-\frac{\Lambda}{3}\rho^2 A)
(1+\frac{J^2}{\rho^2})\biggr]=0\;
\label{eq:dero}
\end{equation}
or
\begin{equation}
\frac{d}{ds}\biggl[
(\frac{d\rho}{ds})^2+(\frac{1}{A}-\frac{\Lambda}{3}\rho^2 A)
(1+\frac{J^2}{\rho^2})\biggr]=0\;
\end{equation}
Integration then yields
\begin{equation}
(\frac{d\rho}{ds})^2+(\frac{1}{A}-\frac{\Lambda}{3}\rho^2 A)
(1+\frac{J^2}{\rho^2})=c_2\;
\end{equation}
Asymptotic behavior shows that $c_2=1$, and thereby
\begin{equation}
(\frac{d\rho}{ds})^2=1-(\frac{1}{A}-\frac{\Lambda}{3}\rho^2 A)
(1+\frac{J^2}{\rho^2})\;
\end{equation}
Eq.~(\ref{eq:fufa}) together with $J=0$ makes Eq.~(\ref{eq:dero}) as
\begin{equation}
\frac{d^2\rho}{ds^2}+\frac{d}{d\rho}(\frac{-M}{\rho}
-\frac{\Lambda}{6}\rho^2)=0\;
\end{equation}
Since the potential field $\Phi$ is defined as
\[
\frac{d^2\rho}{ds^2}+\frac{d\Phi}{d\rho}=0\;
\]
then
\begin{equation}
\Phi=\frac{-M}{\rho}-\frac{\Lambda}{6}\rho^2\;
\end{equation}
which is equivalent to the Schwarzschild-de Sitter potential \cite{rindII77}.
Although the potential at $r=0$ {\it i.e.} $\rho=\alpha MR(t)$ is very large,
but it is finite. It seems likely that the potential field of massive stars
show this behavior and help us to find a physical mechainsm for fixing
$\alpha$. It will be a great success if observing extra high energy phenomena
in AGN's and cosmic rays, provide a lower limit for $\alpha$.\\

\begin{center}
{\bf REMARKS}
\end{center}

A space-time is said to be spherically symmetric, if it admits the group SO(3)
as a group of isometries, with the group orbits spacelike two surfaces. A
coordinate transformation like $r\to r'=r+\alpha M$ which translates the center
of symmetry and thereby breakdowns spherical symmetry does not belong to SO(3).
We have shown that the solutions of spherically symmetric vacuum Einstein
field equations are necessarily neither static nor incomplete.\\

\appendix{\bf Deflection of Light}
\label{sec:deflec}

We would like to derive the deflection of light formula which introduce in
Sec.II. The orbit is described by Eq.~(\ref{eq:defeq}), that is
\begin{equation}
\phi (r)-\phi_\infty = \int_r ^\infty {\biggl[
\frac{A(r)}{D(r)}\biggr] }^{1/2}
{\biggl[ \frac{D(r)}{D(r_o)}\frac{B(r_o)}{B(r)}-1\biggr] }^{-1/2}dr\;
\label{eq:reeq}
\end{equation}
Using for $A(r)$, $B(r)$ and $D(r)$ the Eq.~(\ref{eq:fal}), we notice that
the integral is defined for $r_o>(3-\alpha )M$. The first term of the second
square root is written as
\begin{equation}
\frac{D(r)}{D(r_o)} \frac{B(r_o)}{B(r)}=\biggl[ {r\over r_o} \biggr]^2 \biggl[
\frac{1+\alpha M/r}{1+\alpha M/r_o} \biggr]^3 \biggl[
\frac{1-(2-\alpha )M/r}{1-(2-\alpha )M/r_o} \biggr] ^{-1}\; \label{eq:sear}
\end{equation}
We want to derive deflection angle up to the second order but the expansion
explicity depends on $\alpha $ and $M/r_o$. Since there is no common valid
range of expansion for the involved parameters, we must treat the necessary and
sufficient cases seperately; {\it i.e.} $\alpha <1$ and $\alpha >1$.\\

We consider first $\alpha <1$. Because of the integral singularity condition,
we always have $r_o>(2-\alpha )M$ and therefore with a simple expansion we get
\begin{eqnarray}
\frac{D(r)}{D(r_o)} \frac{B(r_o)}{B(r)}=\biggl[ {r\over r_o} \biggr]^2
\biggl[ 1 & &+(2(1+\alpha )M-{3\alpha^2 M^2 \over r_o})
\left( {1\over r}-{1\over r_o}\right)\nonumber\\
& &+{(2-\alpha )^2 M^2\over r}\left( {1\over r}-{1\over r_o}\right)
+6\alpha M^2\left( {1\over r}-{1\over r_o}\right)^2+\cdots \biggr] \;
\end{eqnarray}
The argument of the second square root in Eq.~(\ref{eq:reeq}) is then
\begin{eqnarray}
\frac{D(r)}{D(r_o)} \frac{B(r_o)}{B(r)}-1=\biggl[ \left( {r\over r_o}
\right) ^2 -1\biggr] \biggl[ 1 & &-{\left( 2(1+\alpha )M-3\alpha^2 M^2/r_o
\right) r\over r_o(r+r_o)}\nonumber\\
& &-{(2-\alpha )^2 M^2\over r_o(r+r_o)}
+{6\alpha M^2(r-r_o)\over r_o^2(r+r_o)}+\cdots \biggr] \;
\end{eqnarray}
so it gives
\begin{eqnarray}
\biggl[ \frac{D(r)}{D(r_o)} \frac{B(r_o)}{B(r)}-1 \biggr]^{-1/2}=& & \biggl[ 
\left( {r\over r_o}\right) ^2-1 \biggr]^{-1/2}\nonumber\\
& & \times \biggl[ 1+{\left(
2(1+\alpha )M-3\alpha^2 M^2/r_o \right) r\over 2r_o(r+r_o)}
+{(2-\alpha )^2 M^2\over 2r_o(r+r_o)}\nonumber\\
& & \;\;\;\;\;\;\;\;\;\;\;\;\;\;\;{} -{3\alpha M^2(r-r_o)\over r_o^2(r+r_o)}
+{3(1+\alpha )^2 M^2 r^2\over 2r_o^2(r+r_o)^2}+\cdots \biggr]\nonumber\\
\end{eqnarray}
The argument of the first square root in Eq.~(\ref{eq:reeq}) and its expansion
is
\begin{eqnarray}
\biggl[ \frac{A(r)}{D(r)} \biggr] ^{1/2}=& & {1\over r} \biggl[
(1+\alpha M/r)(1-(2-\alpha )M/r)\biggr]^{-1/2}\nonumber\\
=& & {1\over r} \biggl[ 1+{(1-\alpha )M\over r}+{(2\alpha^2 -4\alpha +3)
M^2\over 2r^2}+\cdots \biggr]\;
\end{eqnarray}
Inserting these two equations into Eq.~(\ref{eq:reeq}) gives
\begin{eqnarray}
\phi (r_o)-\phi_ \infty =
& & \int_{r_o} ^\infty {dr\over r [({r\over r_o})^2-1]^{1/2}}
\biggl[ 1+{(1-\alpha )M\over r}+{(2\alpha^2 -4\alpha +3)
M^2\over 2r^2}+\cdots \biggr]\nonumber\\
& & \;\;\;\;\;\;\;{} \times \biggl[ 1+{\left( 2(1+\alpha )M-3\alpha^2 M^2/r_o
\right) r\over 2r_o(r+r_o)}+{(2-\alpha )^2 M^2\over 2r_o(r+r_o)}\nonumber\\
& & \;\;\;\;\;\;\;\;\;\;\;\;\;\;\;\;\;\;\;\;\;\;\;\;\;\;\;\;\;\;\;{}
-{3\alpha M^2(r-r_o)\over r_o^2(r+r_o)}
+{3(1+\alpha )^2 M^2 r^2\over 2r_o^2(r+r_o)^2}+ \cdots \biggr]\;
\end{eqnarray}
The integral is elementary, and gives
\begin{equation}
\phi (r_o)-\phi_ \infty ={\pi \over 2}+{2M\over r_o}+
{2M^2\over r_o^2}\biggl[ {15\pi \over 16}-(1+\alpha )\biggr] +\cdots \;
\end{equation}
The deflection of the orbit from a straight line is
\begin{equation}
\Delta \phi =2| \phi (r_o) - \phi_ \infty | -\pi \; \label{eq:eidef}
\end{equation}
Hence to second order in $M/r_o$, the deflection angle is as follows
\begin{equation}
\label{eq:defsm}
\Delta \phi ={4M\over r_o}+{4M^2\over r_o^2} \biggl[ {15\pi \over 16}
-(1+\alpha ) \biggr] +\cdots\;
\end{equation}\\

Next we consider $\alpha >1$. Since in this case we may also have
$r_o<| 2-\alpha |M$ (e.g. for $\alpha >3$ the integral singularity condition
reduces to $r_o>0$), so in order to have a valid expansion we define the
following parameters
\begin{equation}
a\equiv {\alpha M\over 1+\alpha M/r_o}\;\;\;\;\; \& \;\;\;\;\;
b\equiv {(2-\alpha )M\over 1-(2-\alpha )M/r_o}\; \label{eq:pareq}
\end{equation}
and with $\alpha >1$, we always have
\[
a/r_o<1\;\;\;\;\; \& \;\;\;\;\; b/r_o<1\;
\]
So we rewrite Eq.~(\ref{eq:sear}) and expand it as
\begin{eqnarray}
\frac{D(r)}{D(r_o)} \frac{B(r_o)}{B(r)}=& &\biggl[ {r\over r_o} \biggr]^2
\biggl[ 1+a\left( {1\over r}-{1\over r_o}\right) \biggr]^3 \biggl[
1-b\left( {1\over r}-{1\over r_o}\right) \biggr] ^{-1}\nonumber\\
=& & \biggl[ {r\over r_o} \biggr]^2 \biggl[ 1+(3a+b)\left( {1\over r}
-{1\over r_o}\right) +(3a^2+3ab+b^2)\left( {1\over r}-{1\over r_o}\right)^2
+\cdots \biggr]\nonumber\\
\end{eqnarray}
The argument of the second square root in Eq.~(\ref{eq:reeq}) is then
\begin{equation}
\frac{D(r)}{D(r_o)} \frac{B(r_o)}{B(r)}-1= \biggl[
\left( {r\over r_o}\right)^2-1\biggr] \biggl[
1-{(3a+b)r\over r_o(r+r_o)}+{(3a^2+3ab+b^2)(r-r_o)\over r_o^2(r+r_o)}+
\cdots\biggr]\;
\end{equation}
so it gives
\begin{eqnarray}
\biggl[ \frac{D(r)}{D(r_o)} \frac{B(r_o)}{B(r)}-1 \biggr]^{-1/2}= \biggl[
\left( {r\over r_o}\right)^2-1 \biggr]^{-1/2} \biggl[ 1 & &+{(3a+b)r\over
2r_o(r+r_o)}+{3(3a+b)^2 r^2\over 8r_o^2(r+r_o)^2}\;\;\;\;\;\;\;\;\;\;\nonumber\\
& &-{(3a^2+3ab+b^2)(r-r_o)\over 2r_o^2(r+r_o)}+\cdots\biggr]\nonumber\\
\label{eq:secar}
\end{eqnarray}
The argument of the first square root in Eq.~(\ref{eq:reeq}) is
\begin{equation}
\biggl[ \frac{A(r)}{D(r)} \biggr]^{1/2}={1\over r} \biggl[
(1+\alpha M/r)(1-(2-\alpha )M/r)\biggr]^{-1/2}\; \label{eq:fiar}
\end{equation}
In terms of these new parameters, we have the following equalities
\begin{equation}
1+\alpha M/r\equiv (1-a/r_o)^{-1}\biggl[ 1+a\left( {1\over r}
-{1\over r_o}\right)\biggr]\;
\end{equation}
and
\begin{equation}
1-(2-\alpha )M/r\equiv (1+b/r_o)^{-1}\biggl[ 1-b\left( {1\over r}
-{1\over r_o}\right)\biggr]\;
\end{equation}
Inserting these results into Eq.~(\ref{eq:fiar}) and expanding it up to second
order yields
\begin{eqnarray}
\biggl[ \frac{A(r)}{D(r)}\biggr] ^{1/2}=& &{1\over r}\biggl[ (1-a/r_o)
(1+b/r_o) \biggr]^{1/2}\nonumber\\
& & \times \biggl[ 1-{(a-b)\over 2}\left( {1\over r}-{1\over r_o}\right)
+{(3a^2-2ab+3b^2)\over 8}\left( {1\over r}-{1\over r_o}\right)^2
+\cdots\biggr]\nonumber\\
\label{eq:firar}
\end{eqnarray}
Substituting Eq.~(\ref{eq:secar}) and Eq.~(\ref{eq:firar}) into
Eq.~(\ref{eq:reeq}) gives an elementary integral as
\begin{eqnarray}
\phi (r_o)-\phi_ \infty =
& & \int_{r_o} ^\infty {dr\over r[({r\over r_o})^2-1]^{1/2}}
\biggl[ (1-a/r_o)(1+b/r_o) \biggr]^{1/2}\nonumber\\
& & \;\; {} \times\biggl[ 1-{(a-b)\over 2}\left( {1\over r}
-{1\over r_o}\right) +{(3a^2-2ab+3b^2)\over 8}\left(
{1\over r}-{1\over r_o}\right)^2+\cdots\biggr]\nonumber\\
& & \;\; {} \times\biggl[ 1+{(3a+b)r\over 2r_o(r+r_o)}
+{3(3a+b)^2 r^2\over 8r_o^2(r+r_o)^2}
-{(3a^2+3ab+b^2)(r-r_o)\over 2r_o^2(r+r_o)}+\cdots\biggr]\nonumber\\
\end{eqnarray}
which can be easily integrated
\begin{equation}
\phi (r_o)-\phi_ \infty = {\pi \over 2}+{(a+b)\over r_o}-
{(a+b)(a+3b)\over 2r_o^2}+{15\pi (a+b)^2\over 32r_o^2}+\cdots\;
\end{equation}
This gives the deflection angle in terms of the defined parameters as
\begin{equation}
\Delta \phi ={2(a+b)\over r_o}-{(a+b)(a+3b)\over r_o^2}+
{15\pi (a+b)^2\over 16r_o^2}+\cdots\;
\end{equation}

For expanding the parameters, we must notice the region in which the expansion
is valid. For $r_o>\alpha M$ (and consequently $r_o> | 2-\alpha |M$) the
expansions to second order in $M/r_o$ are
\begin{eqnarray}
a/r_o=& & {\alpha M\over r_o}-{\alpha ^2 M^2\over r_o^2}+\cdots\nonumber\\
b/r_o=& & {(2-\alpha )M\over r_o}+{(2-\alpha )^2 M^2\over r_o^2}+\cdots\;
\end{eqnarray}
Hence the Einstein deflection angle up to the second order in this region
becomes
\begin{equation}
\label{eq:defla}
\Delta \phi ={4M\over r_o}+{4M^2\over r_o^2} \biggl[ \frac{15\pi }{16}
-(1+\alpha ) \biggr] +\cdots\;
\end{equation}

The other region we are interested in, is $r_o<| 2-\alpha |M$ (Actually this
region exists if $\alpha >2.5$) which gives the second order expansions as
\begin{eqnarray}
a/r_o=& & 1-{r_o\over \alpha M}+{r_o^2\over \alpha^2 M^2}+\cdots\nonumber\\
b/r_o=& & -1+{r_o\over (\alpha -2)M}-{r_o^2\over (\alpha -2)^2 M^2}+\cdots\;
\end{eqnarray}
Therefore the Einstein deflection angle up to the second order in this
region is
\begin{equation}
\label{eq:defst}
\Delta \phi ={8r_o\over \alpha (\alpha -2)M}+{4r_o^2\over \alpha^2
(\alpha -2)^2M^2} \biggl[ \frac{15\pi }{16}-5(\alpha -1) \biggr] +\cdots\;
\end{equation}\\

\appendix{\bf Precession of Perihelia}
\label{sec:preces}

In order to derive the Precession of Perihelia expression which is used in
Sec.II, we consider a test particle bound in an orbit around a massive object.
The angle swept is given by Eq.~(\ref{eq:preeq}), that is
\begin{eqnarray}
\phi(r)-\phi(r_-)\nonumber\\
=\int_{r_-}^r & & \biggl[
{D(r_-)\left( B^{-1}(r)-B^{-1}(r_-) \right) -D(r_+)\left( B^{-1}(r)
-B^{-1}(r_+) \right) \over D(r_+)D(r_-)\left( B^{-1}(r_+)-B^{-1}(r_-) \right)}
-{1\over D(r)}\biggr]^{-1/2}\nonumber\\
& & \times \biggl[ \frac{A^{1/2}(r)}{D(r)}\biggr] dr\; \label{eq:prean}
\end{eqnarray}
and the orbit precesses in each revolution by an angle
\begin{equation}
\Delta \phi =2| \phi (r_+) - \phi (r_-)| -2\pi \; \label{eq:preor}
\end{equation}

Following Weinberg, we make the argument of the first square root in
Eq.~(\ref{eq:prean}) a quadratic function of $1/D(r)$ which vanishes at
$D(r_\pm )$, so
\begin{eqnarray}
{D(r_-)\left( B^{-1}(r)-B^{-1}(r_-) \right) -D(r_+)\left( B^{-1}(r)
-B^{-1}(r_+)\right) \over D(r_+)D(r_-)\left( B^{-1}(r_+)-B^{-1}(r_-)
\right)}& &\nonumber\\
-{1\over D(r)}=C\left( {1\over D^{1/2}(r_-)}-{1\over D^{1/2}(r)}\right)
& & \left( {1\over D^{1/2}(r)}-{1\over D^{1/2}(r_+)}\right)\nonumber\\
\label{eq:squro}
\end{eqnarray}
The constant $C$ can be determined by letting $r\to \infty $
\begin{equation}
C=\frac{D(r_+)\left( 1-B^{-1}(r_+)\right) -D(r_-)\left( 1-B^{-1}(r_-)\right) }
{D^{1/2}(r_+)D^{1/2}(r_-)\left( B^{-1}(r_+)-B^{-1}(r_-)\right) }\;
\end{equation}
and its expansion up to second order becomes
\begin{equation}
C=1-2M\left( {1\over r_+}+{1\over r_-} \right)
+2\alpha M^2\left( {1\over r_+^2}+{1\over r_-^2}\right) +\cdots\;
\label{eq:cexp}
\end{equation}
Using Eq.~(\ref{eq:squro}) in Eq.~(\ref{eq:prean}) gives then
\begin{equation}
\phi(r)-\phi(r_-)=C^{-1/2} \int_{r_-}^r dr \biggl[ \frac{A^{1/2}(r)}{D(r)}
\biggr] \biggl[ ({1\over D^{1/2}(r_-)}-{1\over D^{1/2}(r)})
( {1\over D^{1/2}(r)}-{1\over D^{1/2}(r_+)}) \biggr]^{-1/2}\;
\end{equation}
Intoducing a new variable $\psi $
\begin{equation}
{1\over D^{1/2}}={1\over 2}\left( {1\over D^{1/2}(r_+)}+{1\over
D^{1/2}(r_-)}\right) +{1\over 2} \left( {1\over D^{1/2}(r_+)}-
{1\over D^{1/2}(r_-)}\right) \sin\psi\;
\end{equation}
makes the integral as
\begin{eqnarray}
\phi(r)-\phi(r_-)=C^{-1/2} \int_{-\pi /2}^\psi d\psi \biggl[ 1 & & -M
\left( {1\over D^{1/2}(r_+)}+{1\over D^{1/2}(r_-)}\right)\nonumber\\  & & -M
\left( {1\over D^{1/2}(r_+)}-{1\over D^{1/2}(r_-)}\right)\sin\psi
\biggr]^{-1/2}\;
\end{eqnarray}
Expanding the square root to second order in $M/D^{1/2}(r)$, and integrating
yields
\begin{eqnarray}
\phi(r_+)-\phi(r_-)=C^{-1/2} \biggl[ \pi & & +{\pi M\over 2}
\left( {1\over D^{1/2}(r_+)}+{1\over D^{1/2}(r_-)}\right)\nonumber\\
& & +{3\pi M^2\over 8}\left( {1\over D^{1/2}(r_+)}
+{1\over D^{1/2}(r_-)}\right)^2 \nonumber\\
& & +{3\pi M^2\over 16}\left( {1\over D^{1/2}(r_+)}
-{1\over D^{1/2}(r_-)}\right)^2+\cdots \biggr]\;
\end{eqnarray}
Using Eq.~(\ref{eq:cexp}) in the above equation and expanding it gives
\begin{eqnarray}
\phi(r_+)-\phi(r_-)=& & \biggl[ 1+M\left( {1\over r_+}+{1\over r_-}\right)
+{3M^2 \over 2}\left( {1\over r_+}+{1\over r_-}\right)^2
-\alpha M^2\left( {1\over r_+^2}+{1\over r_-^2}\right) +\cdots \biggr]\nonumber\\
& & \times \biggl[ \pi +{\pi M\over 2}\left( {1\over r_+}+{1\over r_-}\right)
-{\pi \alpha M^2 \over 2}\left( {1\over r_+^2}+{1\over r_-^2}\right)\nonumber\\
& & \;\;\;\;\;\;\;{}+{3\pi M^2 \over 8}\left( {1\over r_+}+{1\over r_-}\right)^2
+{3\pi M^2 \over 16}\left( {1\over r_+}-{1\over r_-}\right)^2+\cdots \biggr]\;
\end{eqnarray}
So with Eq.~(\ref{eq:preor}), we get the precession per revolution as
\begin{eqnarray}
\Delta \phi=& & 3\pi M\left( {1\over r_+}+{1\over r_-}\right)
+{19\pi M^2 \over 4}\left( {1\over r_+}+{1\over r_-}\right)^2
+{3\pi M^2 \over 8}\left( {1\over r_+}-{1\over r_-}\right)^2\nonumber\\
& & -3\pi \alpha M^2\left( {1\over r_+^2}+{1\over r_-^2}\right) +\cdots\;
\end{eqnarray}
The elements of planetary orbits are ${\it a}$ and ${\it e}$ which defined by
\begin{eqnarray}
r_\pm & & =(1\pm {\it e}){\it a}\nonumber\\
L & & =(1-{\it e}^2){\it a}\;
\end{eqnarray}
where
\begin{equation}
{1\over L}={1\over 2}\left( {1\over r_+}+{1\over r_-}\right)\;
\end{equation}
These result the desired form of precession per revolution up to the second
order as
\begin{equation}
\label{eq:prec}
\Delta \phi=\frac{6\pi MG}{L} \biggl[ 1+\frac{MG}{L}\biggl( \frac{19}{6} +
\frac{{\it e}^2}{4}-2\alpha (1+{\it e}^2) \biggr) \biggr] +\cdots\;
\end{equation}\\

\end{document}